\documentclass[aps,prb,twocolumn,showpacs,superscriptaddress]{revtex4}

\usepackage{graphicx}
\usepackage{dcolumn}
\usepackage{bm}
\usepackage{amsmath}

\begin{document}

\title{Role of remote interfacial phonons in the resistivity of graphene}

\author{Y. G. You}
\affiliation{School of Physics, Konkuk University, Seoul 05029, Korea}
\author{J. H. Ahn}
\affiliation{School of Physics, Konkuk University, Seoul 05029, Korea}
\author{B. H. Park}
\affiliation{School of Physics, Konkuk University, Seoul 05029, Korea}
\author{Y. Kwon}
\affiliation{School of Physics, Konkuk University, Seoul 05029, Korea}
\author{E. E. B. Campbell}

\affiliation{School of Physics, Konkuk University, Seoul 05029, Korea}
\affiliation{EaStCHEM, School of Chemistry, Edinburgh University, David Brewster Road, Edinburgh EH9 3FJ,
United Kingdom}

\author{S. H. Jhang}\altaffiliation{e-mail:shjhang@konkuk.ac.kr}
\affiliation{School of Physics, Konkuk University, Seoul 05029, Korea}

\begin{abstract}
The temperature ($\it T$) dependence of electrical resistivity in graphene has been experimentally investigated between 10 and 400~K for samples prepared on various substrates; HfO$_2$, SiO$_2$ and h-BN. The resistivity of graphene shows a linear $\it T$-dependence at low $\it T$ and becomes superlinear above a substrate-dependent transition temperature. The results are explained by remote interfacial phonon scattering by surface optical phonons at the substrates. The use of an appropriate substrate can lead to a significant improvement in the charge transport of graphene.
\end{abstract}


\maketitle

Graphene is a two-dimensional carbon allotrope that has excellent electrical properties and higher electron mobility compared to silicon~\cite{novoselov2004electric,morozov2008giant,bolotin2008ultrahigh}. These properties make graphene particularly interesting for many applications in electronic devices\cite{novoselov2012roadmap,yang2012graphene}.
Understanding how the substrate affects electron transport is also critical to achieving the promise of intrinsic graphene.
Many experimental studies show the resistivity of graphene is strongly dependent on temperature above $\sim$200~K.
Chen \emph{et~al.}~attributed this to the extrinsic scattering by surface phonons at the substrate\cite{chen2008intrinsic}, suggesting the importance of substrate choice for graphene devices.
Recently, I-Tan\emph{ et~al.}~theoretically investigated the dependence of the surface optical phonon scattering in graphene on various substrates\cite{lin2013surface}.
For HfO$_2$ substrate with lower surface optical phonon energy of $\sim$21~meV, the resistivity arising from the surface optical phonon scattering was calculated to be $\sim$600~$\Omega$ at room temperature for a Fermi energy $E_{\text{F}}$ of 100~meV, much larger than $\sim$10~$\Omega$ calculated for an h-BN substrate\cite{lin2013surface}.
Here we prepare graphene devices on three different substrates of HfO$_2$, SiO$_2$ and h-BN, and  report experimental investigations of the effect of remote interfacial phonons on the resistivity of graphene.

\begin{figure}
\includegraphics[width=3.2in]{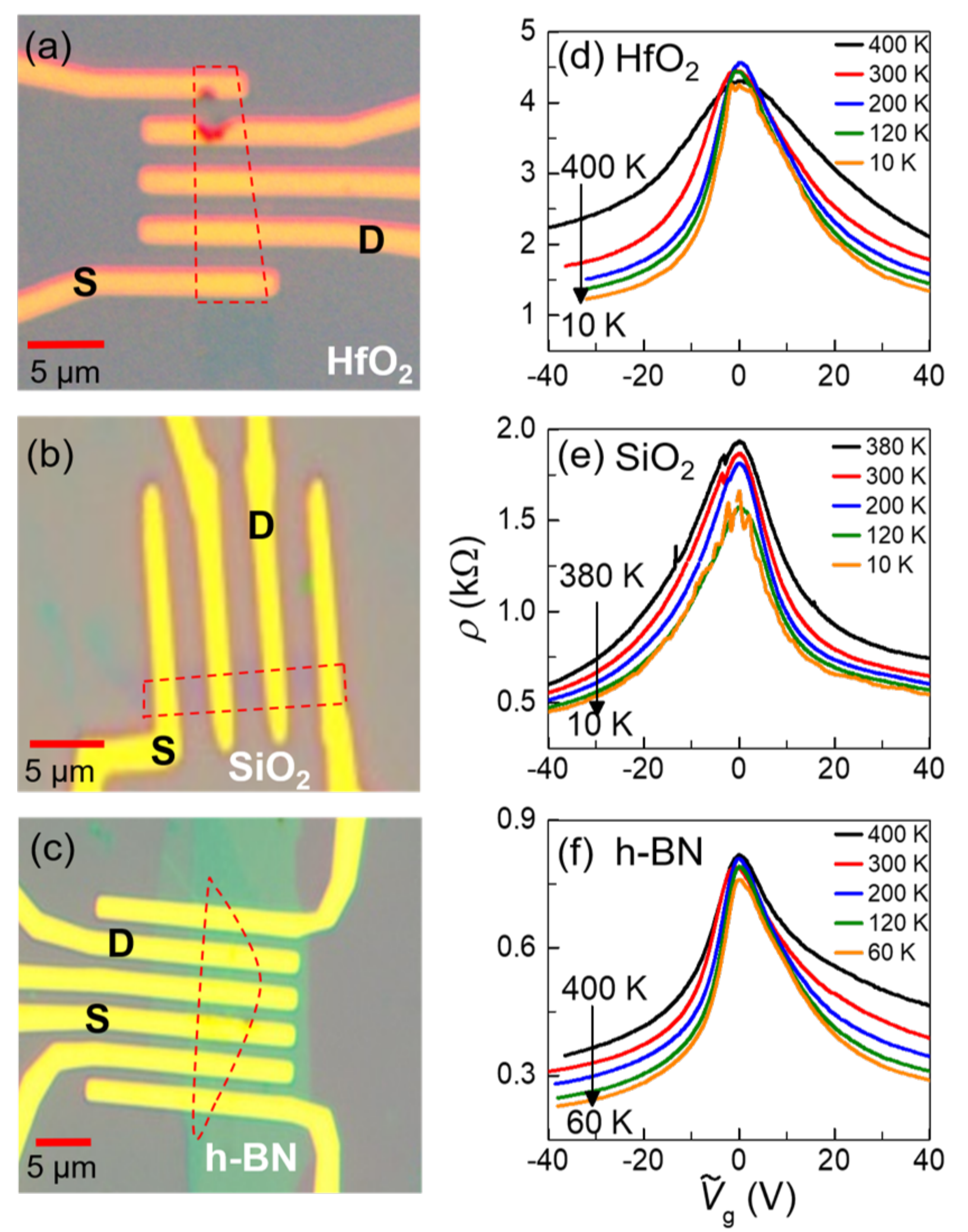}
\caption{(Color online) Optical images of graphene samples prepared on (a) HfO$_2$, (b) SiO$_2$, and (c) h-BN, respectively. Red dashed lines indicate the position of graphene.
S and D represent the source and the drain electrode, used in electrical measurement. Two-probe resistivity versus gate voltage \emph{\~{V}}$_g$, obtained at various fixed temperatures for the graphene on (d) HfO$_2$, (e) SiO$_2$, and (f) h-BN. Charge neutrality point was adjusted to be located at \emph{\~{V}}$_g=0$~V in order to compare the influence of various substrates at similar carrier densities.
}
\end{figure}

Experiments were performed with exfoliated graphene on 300~nm-SiO$_2$/Si substrate \cite{blake2007making}.
Graphene samples were reliably identified as monolayers by means of Raman spectroscopy. \cite{ferrari2006raman,graf2007spatially}
Figs.~1(a), (b) and (c) show optical images of graphene devices fabricated on HfO$_2$, SiO$_2$, and h-BN, respectively.
For the preparation of graphene on HfO$_2$, 30~nm-thick HfO$_2$ was deposited on a SiO$_2$/Si substrate
by using atomic layer deposition (ALD) and graphene was transferred to the HfO$_2$ film.\cite{jiao2008creation}
For the graphene device on h-BN, a 10~nm-thick flake of h-BN was first exfoliated on a SiO$_2$/Si substrate, shown as green in Fig. 1(c), and then the graphene was transferred onto the h-BN flake. Electrodes were defined by using a conventional electron-beam lithography technique, and Pd(20~nm)/Au(30~nm) electrodes were deposited on top of the graphene.
The temperature $T$ dependence of resistivity $\rho$ was studied between 10~and 400~K in a liquid helium system (Quantum Design PPMS).

Figs.~1(d), (e) and (f) display the resistivity as a function of gate voltage \emph{\~{V}}$_g$ at various fixed $T$ for graphene on HfO$_2$, SiO$_2$, and h-BN, respectively.
The resistivity of graphene was larger on the HfO$_2$ substrate for all applied gate voltages.
Near the charge neutrality point (CNP), the resistivity of the graphene on HfO$_2$ was five times larger than that of the graphene on h-BN.
The mobility in graphene, estimated from positive gate voltages at $T = 60$~K, was $\sim$3000, $\sim$5000 and $\sim$17000~cm$^2$/Vs for HfO$_2$, SiO$_2$, and h-BN substrates, respectively.
In the low-$T$ limit, the resistivity saturated at a temperature-independent value of $\rho_0$ for all three substrates, consistent with previous studies of graphene on SiO$_2$.\cite{zou2010deposition,chen2008intrinsic,efetov2010controlling}.
At $n$ = 2 $\times$ 10$^{12}$/cm$^2$, we have $\rho_0$ of 1533, 540, 305 $\Omega$ for HfO$_2$, SiO$_2$, and h-BN, respectively. The contact resistance of the Pd electrodes is about 10 to 30 $\Omega$.\cite{song2012determination,watanabe2012low,xia2011origins}
The residual resistivity $\rho_0$ mainly originates from electron scattering with static impurities and point defects.
The h-BN substrate has an atomically smooth surface with less charge traps and dangling bonds,\cite{dean2010boron} and in addition possesses a larger surface optical phonon energy of $\sim$101~meV, compared to $\sim$59 and $\sim$22~meV of SiO$_2$ and HfO$_2$, respectively. Therefore, the observed higher mobility and the smaller resistivity of the graphene on the h-BN substrate can be expected.

\begin{figure}
\includegraphics[width=3.2in]{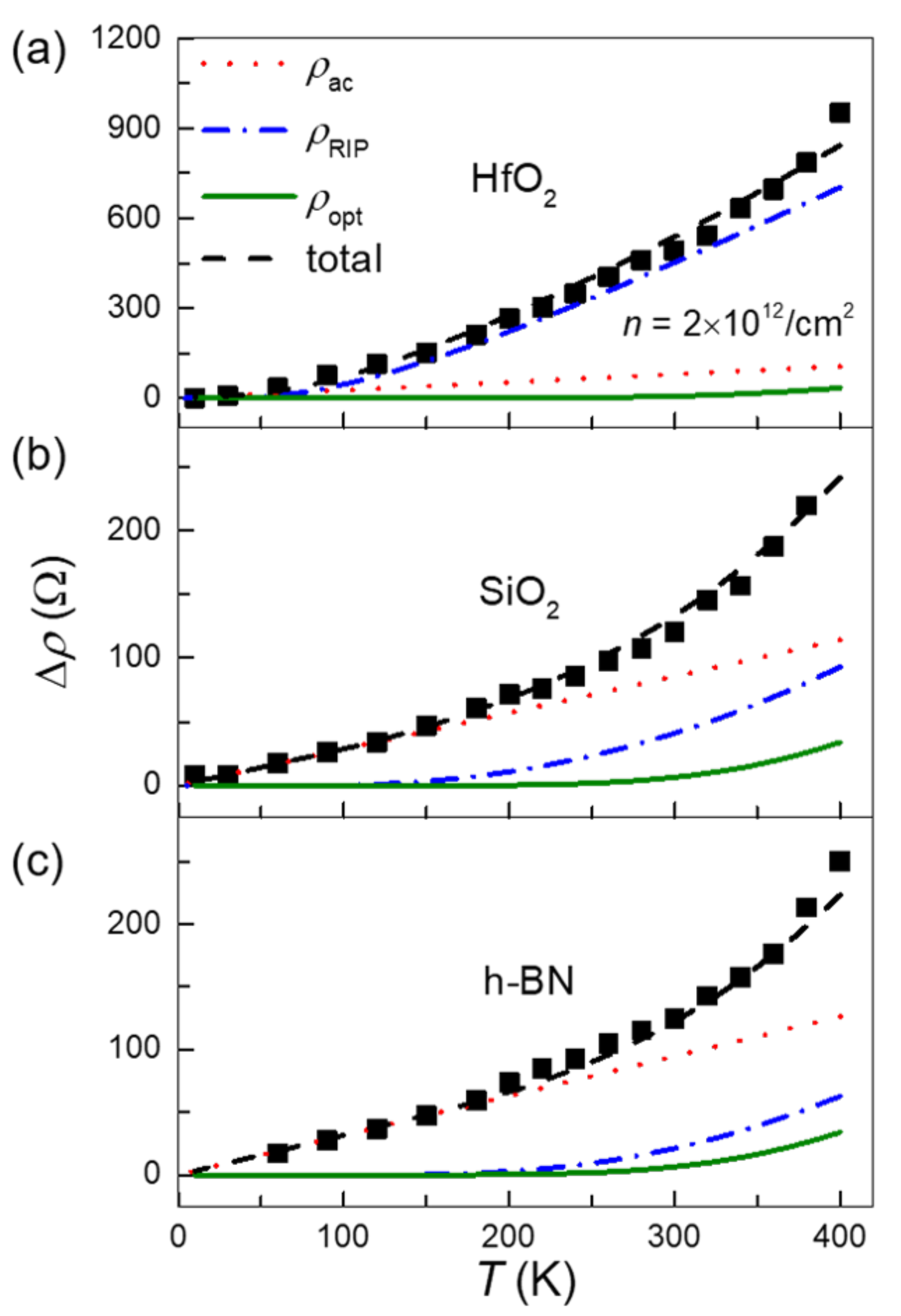}
\caption{(Color online) Temperature dependent part of resistivity $\Delta\rho$, extracted from Fig.~1 in a doped regime at $n$ = 2 $\times$ 10$^{12}$/cm$^2$ ($E_{\text{F}}\sim170$~meV) for graphene sample on (a) HfO$_2$, (b) SiO$_2$, and (c) h-BN substrate, respectively. $\Delta\rho$ was fitted by three different contributions to the resistivity; from the scattering with acoustic phonons ($\rho_{\text{ac}}$, red dotted line), remote interfacial phonons ($\rho_{\text{RIP}}$, blue dashed line), and optical phonons ($\rho_{\text{opt}}$, green solid line).
}
\end{figure}

To investigate the role of surface optical phonons in more detail, we compare only the $T$-dependent part of the resistivity, $\Delta\rho = \rho(T) - \rho_0$, for graphene samples prepared on the three different substrates.
In Fig.~2, we plot $\Delta\rho$, extracted from Fig.~1 for various $T$ at $n$ = 2 $\times$ 10$^{12}$/cm$^2$.
Note $\Delta\rho$ is much larger for the graphene on HfO$_2$.

We analyze and fit the $T$-dependent resistivity $\Delta\rho$ by separating into three different contributions to the resistivity as follows.
\begin{equation}
\label{eq:resistivity}
\Delta\rho (T,n)= \rho_{\text{ac}} (T)  + \rho_{\text{RIP}} (T,n)  + \rho_{\text{opt}} (T)
\end{equation}

$\rho_{\text{ac}}$($\it T$) is the resistivity due to acoustic phonon scattering, and is linearly proportional to $T$.\cite{hwang2008acoustic} It is independent of carrier density $n$ and is given by

\begin{equation}
\label{eq:resistivity-LA}
\rho_{\text{ac}}(T)= {{\pi^2 D_{\text{A}}^2 k_{\text{B}} T}\over{2 e^2 h \rho_{\text{s}} v_{\text{s}}^2 v_{\text{F}}^2 }}.
\end{equation}

Here $k_{\text{B}}$ is Boltzmann's constant, with $e$ and $h$ being the elementary charge and Planck constant, respectively.
$\rho_{\text{s}}$ = 7.6 $\times$ 10$^{-7}$  kg m$^{-2}$ is the mass density of graphene.
$v_{\text{F}}=10^6$~m/s is the Fermi velocity and $v_{\text{s}}=2.1 \times 10^4$~m/s is the speed of sound.
Red dotted lines in Fig.~2 indicate the contribution due to the acoustic phonon scattering and the slopes give the acoustic deformation potential $D_{\text{A}}$= 30$\pm$3~eV, in good agreement with previous reports. \cite{hwang2008acoustic,dean2010boron,efetov2010controlling,bolotin2008temperature}
$\Delta\rho$ shows a linear dependence at lower $T$ and becomes superlinear at higher $T$, implying scattering with high-energy phonon modes.
Note, whereas $\Delta\rho$ follows a linear behavior up to $\sim200$~K for the graphene on h-BN (Fig.~2(c)), it deviates from the linear dependence already well below 100~K for the graphene on HfO$_2$ (Fig.~2(a)).

The second term, $\rho_{\text{RIP}}$($\it T,n$), expresses the resistivity contribution due to the scattering with remote interfacial phonons in the substrates, and can be written as \cite{chen2008intrinsic,lin2013surface,zou2010deposition}

\begin{equation}
\label{eq:resistivity-SO}
\rho_{\text{RIP}} (T,n)= C \left({ 1\over e^{{\hbar\omega_1}/{k_B T}}-1} +{S\over e^{{\hbar\omega_2}/ {k_B T}}-1} \right)
\end{equation}

\begin{table}[t]
\caption{\label{tab:table1}
~Surface optical phonon energy of various substrates}
\begin{ruledtabular}
\begin{tabular}{cccc}
Quantity (units) & HfO$_2$\cite{zou2010deposition,fischetti2001effective} & SiO$_2$\cite{fischetti2001effective,fratini2008substrate} & h-BN\cite{perebeinos2010inelastic,geick1966normal} \\
\hline
$\hbar\omega_1$ (meV) & 21 & 59 & 101\\
$\hbar\omega_2$ (meV) & 54 & 155 & 196 \\
Scattering ratio $S$ & 0.96 & 6.83 & 2.17 \\
\end{tabular}
\end{ruledtabular}
\end{table}

\begin{figure}
\includegraphics[width=3.2in]{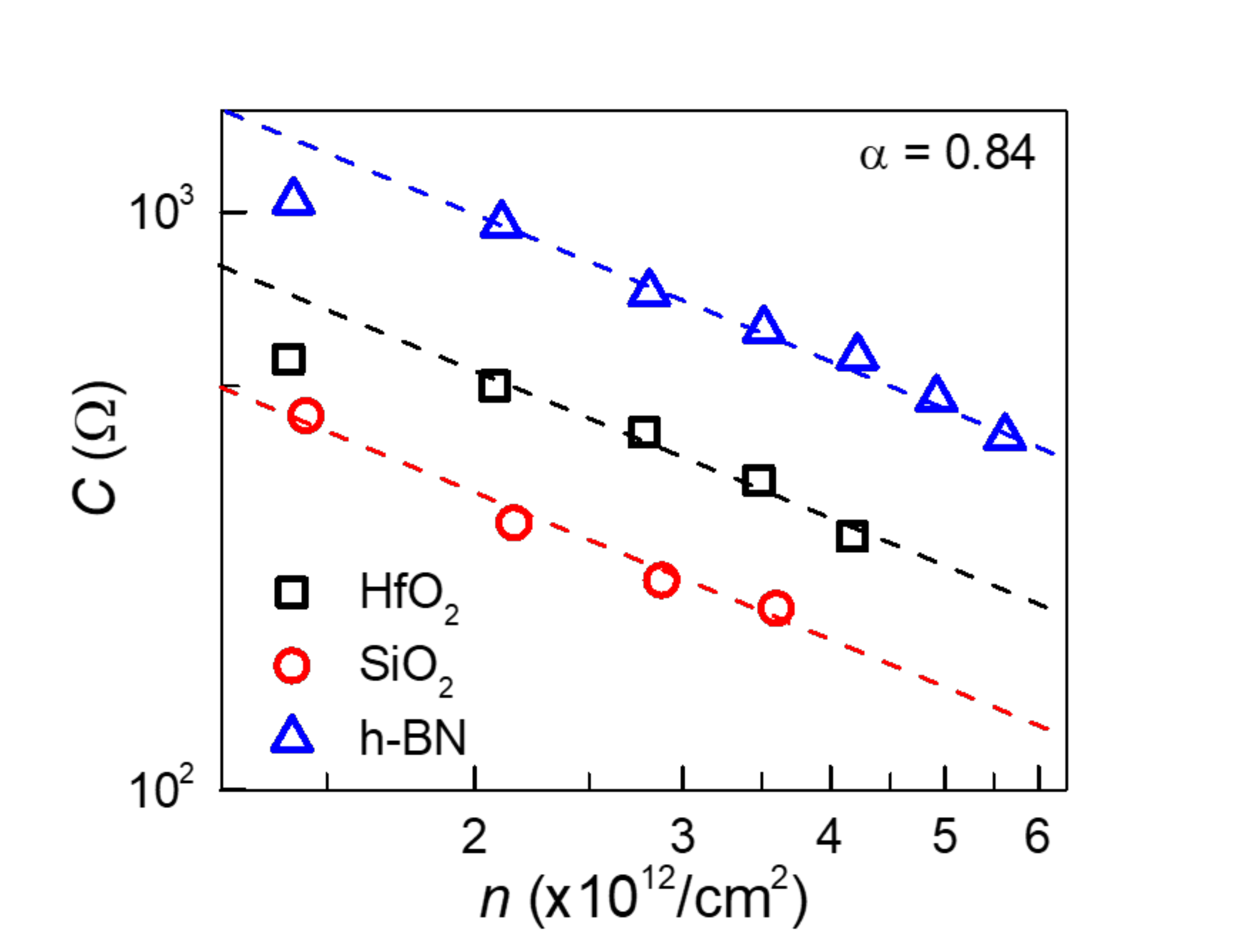}
\caption{(Color online) Resistivity coefficient vs carrier density for each substrate. Dashed lines are fits with $\alpha$ = 0.84.
}
\end{figure}

Here ${\hbar\omega_1}$ and ${\hbar\omega_2}$ represent the energy of the two strongest surface optical phonon modes, with \textit{S} being the ratio of coupling to the electrons.
We used ${\hbar\omega_1}\approx21$~meV and ${\hbar\omega_2}\approx$~54~meV for HfO$_2$, 59 and 155~meV for SiO$_2$, and 101 and 196~meV for the h-BN substrate, respectively.
In table I, the values of ${\hbar\omega_1}$, ${\hbar\omega_2}$ and $S$ are summarized for the three different substrates.\cite{zou2010deposition,perebeinos2010inelastic,geick1966normal,fischetti2001effective,fratini2008substrate}
$C = \textit{BV}_g^{-\alpha}$ is a resistivity coefficient derived from fitting parameters $\it{B}$ and $\alpha$.\cite{chen2008intrinsic}

In Fig.~3, the resistivity coefficient $\it{C}$ is plotted as a function of the charge carrier density. The dashed lines represent fits with $\alpha$ = 0.84.
The resistivity coefficient decreases as carrier density increases due to the charge carrier screening effect.\cite{adam2007self,siegel2013charge}
The difference in $\it{C}$ at the same carrier density is partly related to the difference in effective distance $\it{d}$ between the graphene and the substrate for each device.
Following the approach of Ref.~\onlinecite{lin2013surface}, and using the relation, $B = {{1.84\times10^{-9}hF_1^2}/{de^2(\sqrt{2}+a)^2}}$,
we obtain $\it{d}\simeq$ 4.3, 1.4 and 1.3 \AA\ for the graphene samples on HfO$_2$, SiO$_2$, and h-BN, respectively.
Here, $\it{F_\text{1}}$ is the coupling parameter for the surface optical phonon mode and $a = {{e^2}/({\varepsilon_{avg}\pi\hbar {v_F})}}$,
with $\varepsilon_{\text{avg}}$ being the average dielectric permittivity of the air/graphene/substrate.\cite{lin2013surface}
Larger effective distance between the graphene and the HfO$_2$ substrate is attributed to the roughness of the HfO$_2$ after the ALD growth.
An atomic force microscope (AFM) study shows much larger root-mean-square surface roughness of 10-14 \AA\ for our HfO$_2$ substrates, compared to $\sim$5 and 1-2 \AA\ for SiO$_2$, and h-BN, respectively.

The last term, $\rho_{\text{opt}}$~($\it T$) is the resistivity component due to electron scattering with the optical phonons of graphene.
A$_1^\prime$ phonon mode at K point, longitudinal (LO) and transverse optical (TO) phonon modes at $\Gamma$ point are considered. The A$_1^\prime$ mode has an energy of $\sim$150~meV, and LO and TO modes have an energy of $\sim$200~meV. Owing to the relatively large energy, $\rho_{\text{opt}}$ becomes important only above 300~K (Fig.~2). The resistivity induced by scattering between electrons and phonons can be expressed as follows.\cite{sohier2014phonon}

\begin{equation}
\label{eq:resistivity-graphene-general}
{1\over \rho} = {{e^2 v_F^2}\over 2} \int d\epsilon  DOS(\epsilon)\tau(\epsilon)\left( - {{\partial f^{(0)}} \over {\partial \epsilon }}{(\epsilon)}\right)
\end{equation}

$\tau(\epsilon)$ is the relaxation time and depends on the optical phonon mode. $f^{(0)}$ is the equilibrium Fermi-Dirac distribution function. In the above equation, the density of states (DOS) of graphene and the relaxation time of each mode are substituted to calculate the resistivity component due to optical phonons.

\begin{figure}
\includegraphics[width=3.2in]{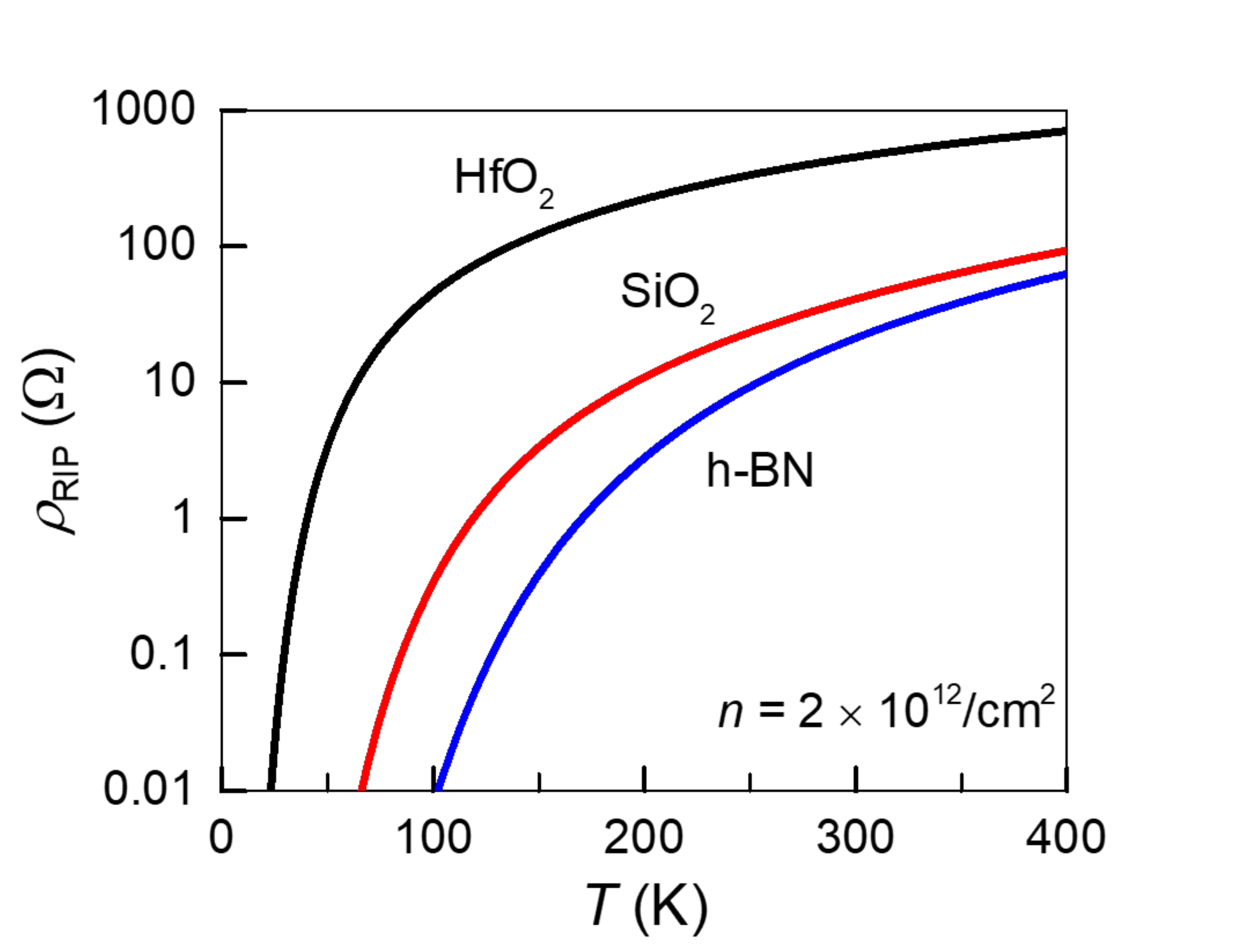}
\caption{(Color online) Resistivity due to the remote interfacial phonon scattering as a function of temperature at $n$ = 2 $\times$ 10$^{12}$/cm$^2$ ($E_F$ $\sim$ 170 meV).
}
\end{figure}

\begin{figure}
\includegraphics[width=3.2in]{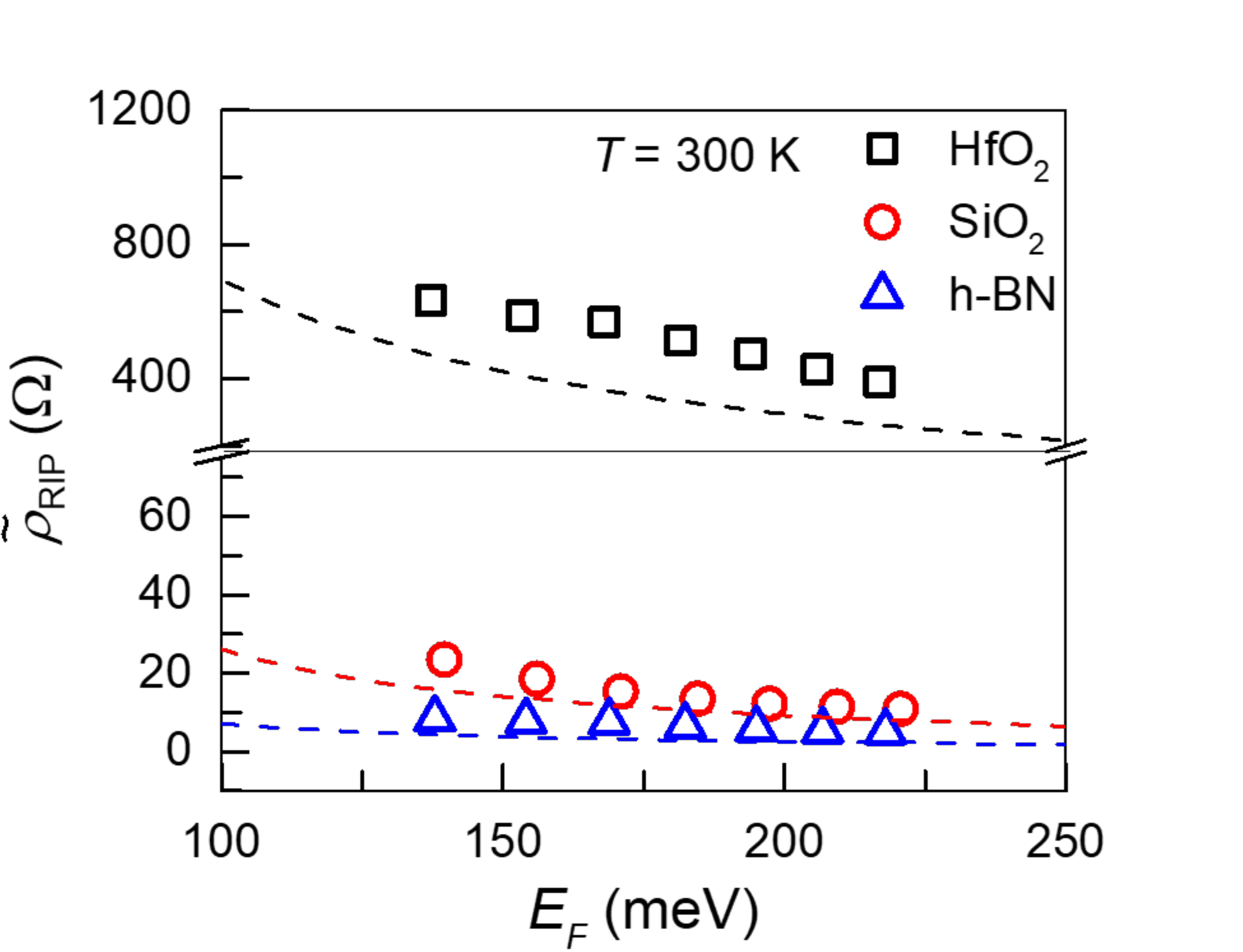}
\caption{(Color online) $\tilde{\rho}_\text{RIP}$, normalized to the effective graphene-substrate distance of 3.4 \AA, shown as a function of Fermi energy at $T=$ 300~K. Dash lines are the theoretical values calculated in the Ref.~\onlinecite{lin2013surface} for the same condition.
}
\end{figure}

In Fig.~4, we plot $\rho_{\text{RIP}}$ as a function of temperature for $n$ = 2 $\times$ 10$^{12}$/cm$^2$ ($E_F\sim$~170~meV).
The resistivity arising from the surface phonon scattering is much larger for graphene on the HfO$_2$ substrate. $\rho_{\text{RIP}}$ becomes larger than 10~$\Omega$ above 60~K for HfO$_2$, significantly at lower $T$, compared to 200~and 250~K for SiO$_2$ and h-BN, respectively. At $T= 300~$K, it reaches $\sim$450~$\Omega$ for graphene on the HfO$_2$, whereas it remains below $\sim$20~$\Omega$ for the h-BN substrate.
Surface optical phonon energies of HfO$_2$, SiO$_2$ and h-BN are 21, 59 and 101~meV, respectively. The smaller surface phonon energy for HfO$_2$ explains in general our results,
yet larger $d\sim4.3$~\AA~ for graphene on HfO$_2$ leads to the underestimation of $\rho_{\text{RIP}}$ against values from other two substrates with $d\sim1.3$~\AA.
Note $\rho_{\text{RIP}}$ is inversely proportional to the effective graphene-substrate distance.

To compare our data with theoretical values from Ref.~\onlinecite{lin2013surface}, calculated for $d=3.4$~\AA,
we normalize $\rho_{\text{RIP}}$ to the effective graphene-substrate distance of 3.4~\AA~and display it in Fig.~5 as a function of the Fermi energy at $T =$ 300~K.
$\tilde{\rho}_\text{RIP}$ decreases as the Fermi energy increases because of the charge carrier screening effect\cite{adam2007self,siegel2013charge}.
Our data are well matched to the theoretically calculated values, verifying the role of the remote interfacial phonons.
Concerning the slight mismatch between our data and the theoretical values, we ignored in our discussion the out-of-plane (ZO) phonon mode in graphene.
At ripples or bubbles, graphene floats locally and the out-of-plane ZO phonon mode ($\sim100$~meV) may play a role.\cite{morozov2008giant,castro2010limits}
Also, we neglected the temperature dependence of contact resistance. A linear $T$-dependence of contact resistance was reported for Pd-graphene contact.\cite{xia2011origins}
This can result in the overestimation of the acoustic deformation potential $D_{\text{A}}$ by $\sim3$~eV in our previous discussion. On the other hand, the non-linear $T$-dependent term for contact resistance is negligible.

In summary, we have studied the effect of surface optical phonons on the resistivity of graphene, prepared on three different substrates, HfO$_2$, SiO$_2$, and h-BN.
For the h-BN substrate, with the larger surface optical phonon energy, the acoustic phonon scattering dominates and the resistivity shows a linear $T$-dependence up to 200~K and then becomes superlinear.
For the HfO$_2$ substrate, with the lower surface optical phonon energy, the remote interfacial phonon scattering strongly dominates and results in much larger resistivity at room temperature.
Our study suggests the use of an appropriate substrate can lead to a significant improvement in the charge transport of graphene.

\begin{acknowledgments}
This paper was supported by Konkuk University in 2015.
\end{acknowledgments}

\end{document}